\documentclass[usegraphicx]{mn2e}
\usepackage{times}

\newlength{\colwidth}
\title[NIR spectroscopy of submillimetre galaxies]{Deep
near-infrared spectroscopy of submillimetre-selected galaxies}
\author[C.\ Simpson et al.]{Chris Simpson$^1$\thanks{E-mail:
chris.simpson@durham.ac.uk}, J.\ S.\ Dunlop$^2$,
S.\ A.\ Eales$^3$, R.\ J.\ Ivison$^{2,4}$, S.\ E.\ Scott$^2$, S.\ J. Lilly$^5$
\newauthor and T. M. A. Webb$^6$\\
$^1$Department of Physics, University of Durham, South Road, Durham
DH1~3LE\\
$^2$Institute for Astronomy, University of Edinburgh, Royal Observatory,
Blackford Hill, Edinburgh EH9~3HJ\\
$^3$Department of Physics and Astronomy, Cardiff University, PO Box 913,
Cardiff CF2~3YB\\
$^4$Astronomy Technology Centre, Royal Observatory, Blackford Hill,
Edinburgh EH9~3HJ\\
$^6$Institut f\"{u}r Astronomie, ETH H\"{o}nggerberg, HPF~G4.1, CH-8093
Zurich, Switzerland\\
$^5$Sterrewacht Leiden, Neils Bohrweg 2, Leiden 233CA, Netherlands}
\begin{document}

\date{Version of \today}

\pagerange{\pageref{firstpage}--\pageref{lastpage}} \pubyear{2003}

\maketitle

\label{firstpage}

\begin{abstract}
We present the results of deep near-infrared spectroscopy of seven
submillimetre-selected galaxies from the SCUBA 8-mJy and CUDSS
surveys. These galaxies were selected because they are too faint to be
accessible to optical spectrographs on large telescopes. We obtain a
spectroscopic redshift for one object, and likely redshifts for two more,
based on a combination of marginal emission line detections and the shape
of the continuum. All three redshifts broadly agree with estimates from
their radio/submm spectral energy distributions. From the emission line
strengths of these objects, we infer star formation rates of
10--25\,M$_\odot\rm\,yr^{-1}$, while the lack of detections in the other
objects imply even lower rates. By comparing our results with those of
other authors, we conclude it is likely that the vast majority (more than
90\,per cent) of the star formation in these objects is completely
extinguished at rest-frame optical wavelengths, and the emission lines
originate in a relatively unobscured region. Finally, we look at future
prospects for making spectroscopic redshift determinations of submm
galaxies.
\end{abstract}


\section{Introduction}

The resolution of the majority of the extragalactic submillimetre (submm)
background into individual sources (Blain et al.\ 1999) has made it
possible, in principle, to study the redshift distribution of these sources
and hence investigate the history of obscured star formation in the
Universe. Such investigations are important to compare the star formation
which occurs in obscured sources with that which takes place apparently
unobscured, e.g., in the Lyman break galaxies identified by Steidel,
Pettini \& Hamilton (1995). Although the similar strengths of the
extragalactic optical and submm backgrounds suggest that obscured and
unobscured star formation occur in roughly equal quantities (Hauser et al.\
1998), the more detailed relationship between the populations of galaxies
within which these processes occur is unclear. Are they coeval populations,
or are each of them the dominant mode of star formation at a particular
cosmic epoch?

Spectroscopy of submm-selected sources has proven difficult and
time-consuming, and has led to the emergence of an industry to
estimate redshifts using template SEDs which cover the wavelength
range from the far-infrared to the radio. Since the initial work by
Carilli \& Yun (1999, 2000), the technique has been adapted by Dunne,
Clements \& Eales (2000), Rengarajan \& Takeuchi (2001), Yun \&
Carilli (2002), and Hughes et al.\ (2002). Often these estimates
disagree strongly, however (see, e.g., Section~5 of this paper), due
to differences in the assumed spectral energy distribution templates,
and this only increases the need for spectroscopic measurements.

There are two main issues which make it difficult to obtain spectroscopy of
these objects. The first of these arises because the spatial resolution of
the Submillimetre Common User Bolometer Array (SCUBA) is rather poor
($15''$ FWHM), leading to a number of optical/infrared counterparts within
the source error circle. The case of HDF~850.1, the brightest submm source
in the Hubble Deep Field, serves as a good example (Hughes et al.\ 1998;
Downes et al.\ 1999; Dunlop et al.\ 2004). To obtain an accurate position
requires sensitive observations with an interferometer at millimetre or
centimetre wavelengths in order to achieve the arcsecond resolution
necessary to make an unambiguous identification.

The second difficulty arises after an accurate position has been
determined. In many cases, the correct identification is extremely faint
and may require deep imaging observations to be identified.  Ivison et al.\
(2002; hereafter I02) report the optical/IR continuum characteristics of a
relatively bright submm sample, having refined most of their positions via
deep radio imaging. A large fraction was shown to be sufficiently bright
for spectroscopy on large telescopes, as later demonstrated using the
LRIS-B UV/optical spectrograph on the Keck telescope (Chapman et al.\
2003). The tendency for the optical/IR counterparts to be diffuse and/or
multiple (``composite red/blue galaxies''; I02) means that the fraction of
light transmitted by a spectroscopic slit is low compared to, say, a
Lyman-break galaxy of similar magnitude. Continuum magnitudes are not a
reliable indicator of likely success, however, if targets display strong
emission lines, as many have been shown to do, including several with $I
\ge 26$ (we use Vega magnitudes throughout). Nevertheless, the ability to
detect continuum emission with a reasonable signal-to-noise ratio is
essential if emission lines are weak or lie outside the available
wavelength coverage, and conclusions about the underlying stellar
population can only be drawn if there is sufficient signal-to-noise ratio
in the continuum to detect breaks and/or stellar absorption feautures.
Given the optical faintness and frequent red colours of these sources,
sensitive near-infrared spectroscopy may prove to be a fruitful avenue for
sources where optical spectroscopy has proven unsuccessful.

In this paper, we present the results from deep near-IR spectroscopy
of a sample of submm galaxies, made with the extremely sensitive OHS
instrument on Subaru Telescope. The format is as follows. In
Section~2, we describe the selection of our sample and our
observations, and in Section~3, we detail the careful data reduction
procedures required to extract maximum sensitivity from these faint
objects. Section~4 presents our results, while in Section~5 we
describe the spectra of individual objects in detail. We discuss our
results in Section~6, and speculate on future prospects for obtaining
complete spectroscopic redshifts for submillimetre samples in
Section~7. Finally, we summarize our results and conclusions in
Section~8. Throughout this paper, we adopt $H_0 =
70\rm\,km\,s^{-1}\,Mpc^{-1}$, $\Omega_{\mathrm m}=0.3$, and
$\Omega_\Lambda=0.7$.

\section{Sample selection and observations}

Our sample of targets was taken from the SCUBA 8-mJy Survey of the Lockman
Hole East and ELAIS~N2 fields (Scott et al.\ 2002; Fox et al.\ 2002) and
the Canada--UK Deep Submillimetre Survey (CUDSS) 14-hour field (Eales et
al.\ 2000; Webb et al.\ 2003). We selected only targets which had been
securely identified through follow-up radio observations which were able to
provide sufficient astrometric accuracy to isolate a single
object\footnote{ELAIS~N2~850.12 was not formally detected in the radio,
though I02 note that the combination of its red colour and the presence of
several 3$\sigma$ radio peaks makes the identification reasonably
secure.}. Furthermore, we selected targets for which the identification had
very red optical/IR colours (see I02), so as to exclude galaxies for which
optical spectroscopy might be expected to be more successful. The sample is
therefore by no means complete, and was constructed to test the limits of
the most sensitive near-infrared spectrograph currently available on any
telescope and assess the likely success of future IR spectroscopy.

Galaxies were observed with the OH-airglow suppressor (OHS; Iwamuro et al.\
2001) and the Cooled Infrared Spectrograph and Camera for OHS (CISCO;
Motohara et al.\ 2002) on Subaru Telescope during the period UT 2002 May
20--24. OHS is a spectroscopic filter which eliminates 224 strong OH
airglow lines in the $J$ and $H$ bands from a $1.3 \times 28\rm\,arcsec^2$
region of sky and outputs an undispersed beam which is then fed to CISCO to
enable images or spectra to be taken. The broad-band sky background in this
suppressed region (the `dark lane') is reduced by a factor of $\sim 25$,
enabling the acquisition of extremely faint targets. Conditions were clear
for all the observations presented here, and the optical seeing was
measured to be 0.5--0.7\,arcsec from the auto-guider camera.

Each target was observed in an identical manner. First, a nearby offset
star was acquired and centred in the OH-suppressed `dark lane', using CISCO
in imaging mode with an $H$-band filter. The telescope was then moved
according to the offset between this star and the target, as measured from
existing near-IR images. The location of the target on the array was then
determined by nodding the telescope along the slit and taking images at
each position, the number and length of which depended on the expected
magnitude of the target. The images were combined to remove the sky
background and allow the target to be seen. If the target was not within
the dark lane, an appropriate telescope offset was applied and the process
repeated until an accurate centroid could be determined. A small offset was
then applied using the autoguider camera to move the target into the centre
of the dark lane. The spectrograph slit was then closed to 1\,arcsec and
aligned with the dark lane, and the {\it JH\/} grism inserted into the
light path. Eight exposures were then taken in two quads using the standard
`ABBA' technique (e.g.\ Eales \& Rawlings 1993), with exposure times of 900
or 1000\,seconds per position. The offset between positions was usually
10\,arcsec, although we chose to align the slit to include any nearby
objects and a smaller offset was used if the the standard value would cause
confusion between the beams. In some cases, a small offset along the slit
was applied between the two quads. When these exposures had finished, the
telescope was slewed to a nearby atmospheric ratioing standard which was
acquired without opening the slit by making small telescope movements
perpendicular to the slit position angle until the count rate from the star
was sufficiently high. The stars used were HIP~52256, HIP~71078, and
HIP~80419 (Perryman et al.\ 1997) for the Lockman Hole, CUDSS~14-h, and
ELAIS~N2 fields, respectively. Eight exposures (of 5--10\,seconds each)
were taken in a single quad with two frames per position.

Observations were also made of argon and neon calibration lamps and the
planetary nebula NGC~7027, to provide wavelength calibration and to measure
the spectral resolution of the instrument. These observations were made
with a 0.5-arcsec slit to more allow a more accurate determination of the
locations of the emission lines. No shifts were found in the location of
any spectral features, or in the locations of the unsuppressed night-sky
lines, during the course of the run. An observing log is presented in
Table~\ref{tab:obslog}.

\begin{table}
\centering
\caption{Observing log. The dates refer to the spectroscopic
observations. Imaging observations were made immediately prior to the
spectroscopy, except for LE~850.1, which was imaged on UT 2002 May
21 (see Section 5.1).\label{tab:obslog}}
\begin{tabular}{lcrrr}
\hline
& & PA & \multicolumn{2}{c}{Exposure (s)} \\
Name & UT Date & ($^\circ$) & Imaging & Spectroscopy \\
\hline
LE~850.1 & 2002 May 23 & 167 & $12\times50$ & $8\times1000$ \\
LE~850.3 & 2002 May 20 &  43 & $6\times20$ & $8\times1000$ \\
CUDSS~14.1 & 2002 May 20 &  90 & $6\times30$ & $8\times1000$ \\
CUDSS~14.3 & 2002 May 24 & 118 & $6\times50$ & $8\times 900$ \\
CUDSS~14.9 & 2002 May 21 &  63 & $6\times20$ & $8\times 900$ \\
ELAIS~N2~850.2 & 2002 May 23 &  30 & $6\times50$ & $8\times1000$ \\
ELAIS~N2~850.12 & 2002 May 24 &  82 & $6\times50$ & $8\times 900$ \\
\hline
\end{tabular}
\end{table}

\section{Data reduction}

The low level of signal coming from the extremely faint targets (we stress
that our targets are producing detector signals of about 1 electron per
pixel per \textit{minute\/}, which is less than the detector dark current)
makes it imperative to remove all detector artefacts. To do this, we first
subtracted one image from each pair from the other to remove the
contribution from the dark current. Next, any differences in bias level
between the four quadrants of the detector were removed by determining the
signal in unexposed regions of each quadrant and subtracting this constant
value from all the pixels in that quadrant. Variance images for each
combined pair were constructed based on the raw pixel values in the
original images and the detector read noise. Residual sky lines were
removed by fitting a low-order function to the columns of the images,
excluding any regions contaminated by objects. The two pairs in each quad
were then combined, by a straight average if the pixel values in each pair
were consistent at the $5\sigma$ level, or by the value closest to zero if
they were not. Next, the image was shifted by an amount equal to the offset
between the positive and negative beams and this shifted image was
subtracted from the original, with further rejection of deviant pixels.
The variance images were combined in a statistically-appropriate manner.
Then the two quads were combined and spectra of all objects extracted in
1-arcsec (9-pixel) apertures. The pixel-by-pixel uncertainty on this
spectrum was determined from the variance image, although this is likely to
underestimate the true noise level in regions of strong sky emission where
the sky level may not be accurately determined.

Spectra of the atmospheric ratioing standards were obtained by simply
averaging all the spectral images taken in the `A' position and
subtracting the average of the spectral images taken in the `B' position.
This image was shifted to align the two beams and subtracted from the
original version of itself. A spectrum was extracted in a 1-arcsec
aperture.

A quadratic wavelength solution was determined from the arc and planetary
nebula spectra and applied to all the spectra. The r.m.s.\ deviation of the
individual emission lines from the fit was $\sim 9$\,\AA\ (approximately 1
pixel). Model spectra of the standard stars were constructed assuming they
could be described by blackbodies with effective temperature and $H$-band
magnitudes determined from the $V$-band magnitudes and spectral types
listed in the HIPPARCOS catalogue (Tokunaga 2000). Spectra of the targets
in flux units were then determined by dividing each target spectrum by the
appropriate atmospheric standard spectrum and multiplying by the model
blackbody. Since the standard stars were chosen to be close in the sky to
the targets, their airmasses at the time of observation are similar to
those of the target at the end of its observation, rather than an average
target airmass. This may result in imperfect removal of atmospheric
absorption features.

The absolute flux calibration of the targets was determined from the
acuqisition imaging. Images at each of the two positions were combined and
then subtracted from each other. This image was then shifted to align the
two beams and subtracted from the original version of itself. Photometry
was performed in a 1-arcsec square aperture and therefore the flux scale of
the final spectra represents the amount of light in the extraction
aperture. The uncertainty in the photometry was estimated by fitting
separate Gaussians to sigma-clipped, logarithmically-scaled histogram of
the pixel values both inside and outside the dark lane. The instrumental
zeropoint of the photometry was tied to observations of the UKIRT Faint
Standards FS~27 and FS~146 (Hawarden et al.\ 2001), taken at the end of the
last night. Any systematic uncertainties in the absolute calibration
introduced by only taking such images at one point during the run are
likely to be dwarfed by the 10--20\,per cent random uncertainties in the
target photometry. In cases where more than one object was in the
spectrograph slit, the flux scale was tied to the brightest object.

Inevitably, we expect some spurious features from `warm' pixels which are
not rejected by our rejection algorithm, since this algorithm needs to be
conservative to avoid throwing out a large amount of good data (in
addition, the rejection limits need to be loose enough to not flag a change
in the signal which is the result of changes in the seeing and/or
atmospheric transparency). The reduced two-dimensional spectra are
therefore used to determine the veracity of any putative emission lines, by
determining whether they arise from only one of the two beams.

\begin{figure}
\includegraphics[angle=-90,width=\colwidth]{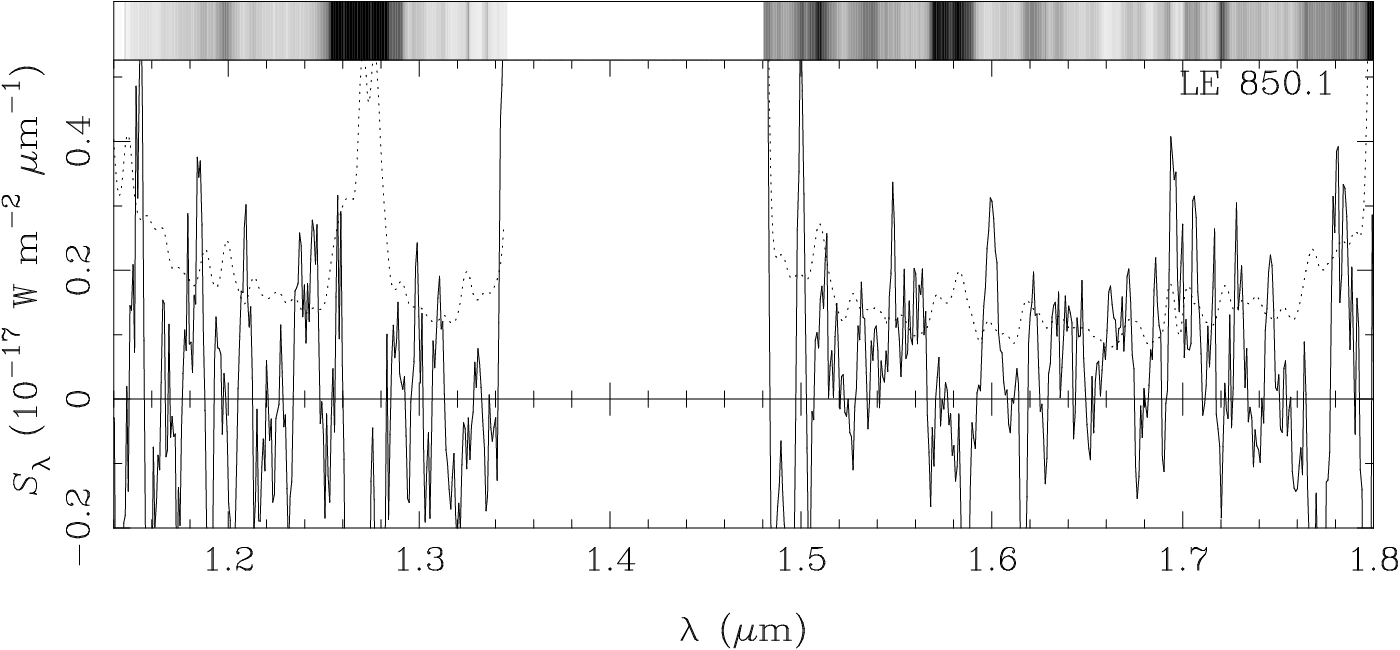}
\includegraphics[angle=-90,width=\colwidth]{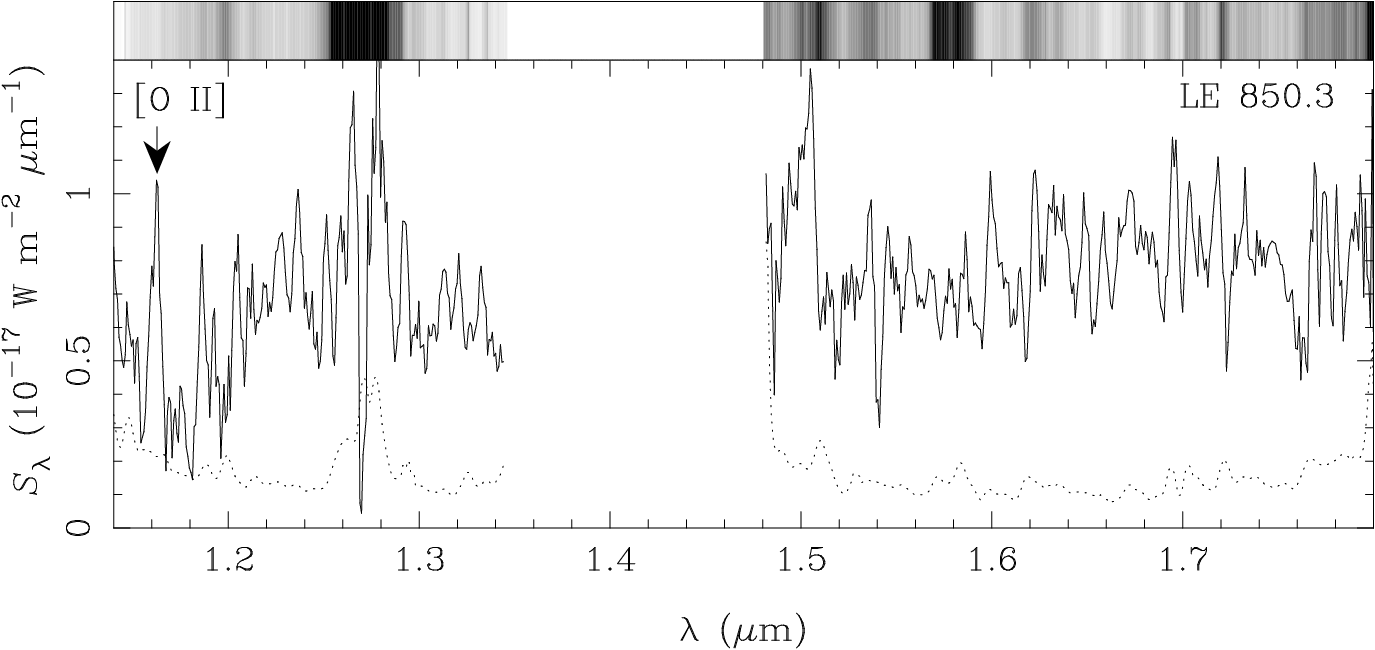}
\includegraphics[angle=-90,width=\colwidth]{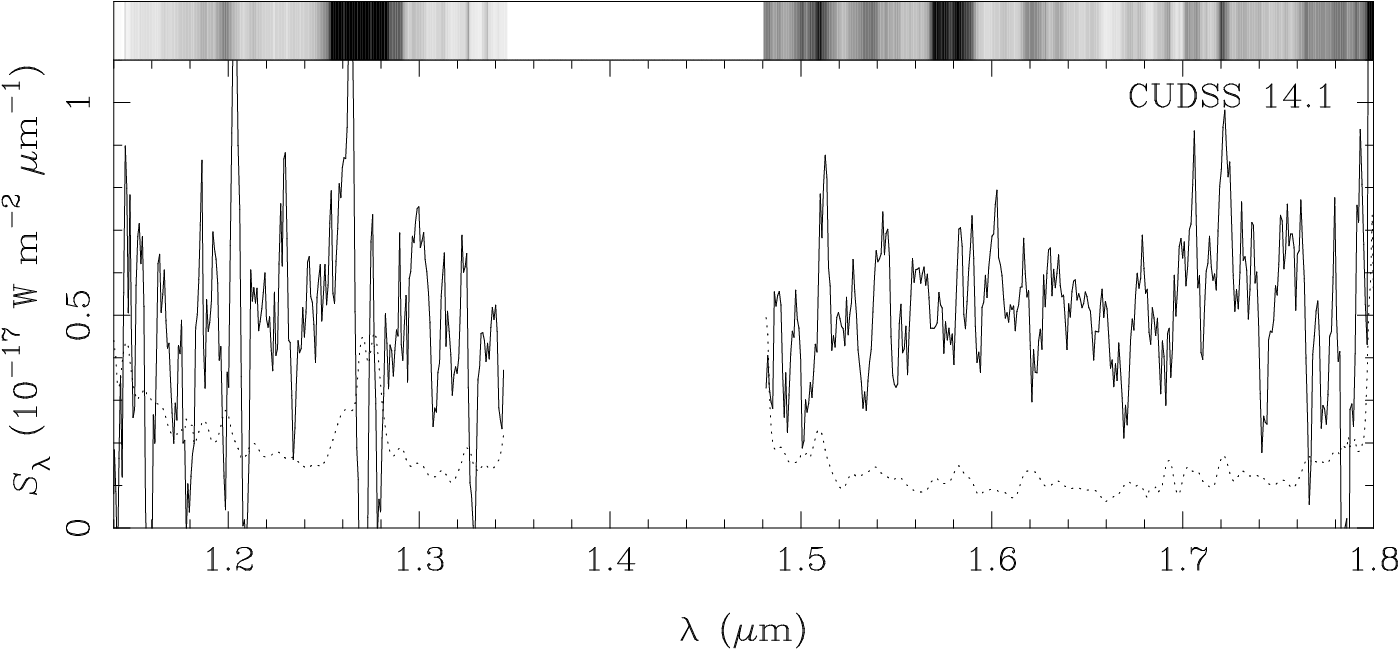}
\includegraphics[angle=-90,width=\colwidth]{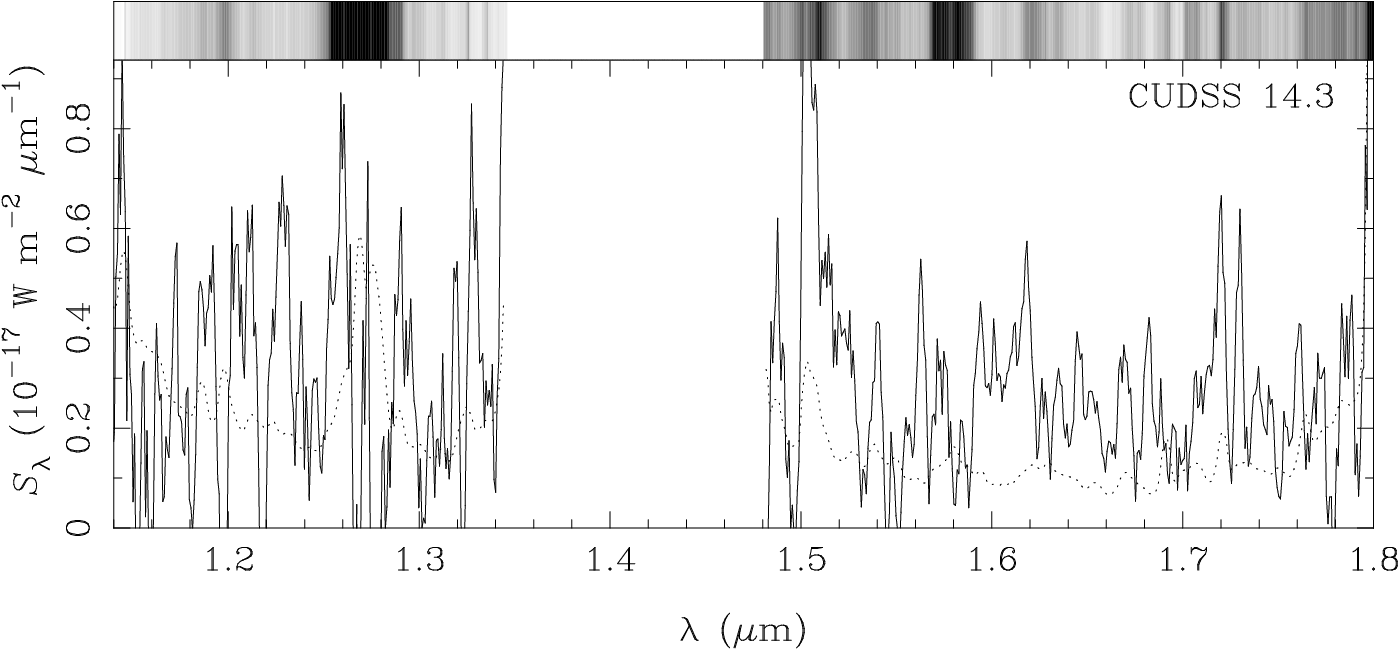}
\includegraphics[angle=-90,width=\colwidth]{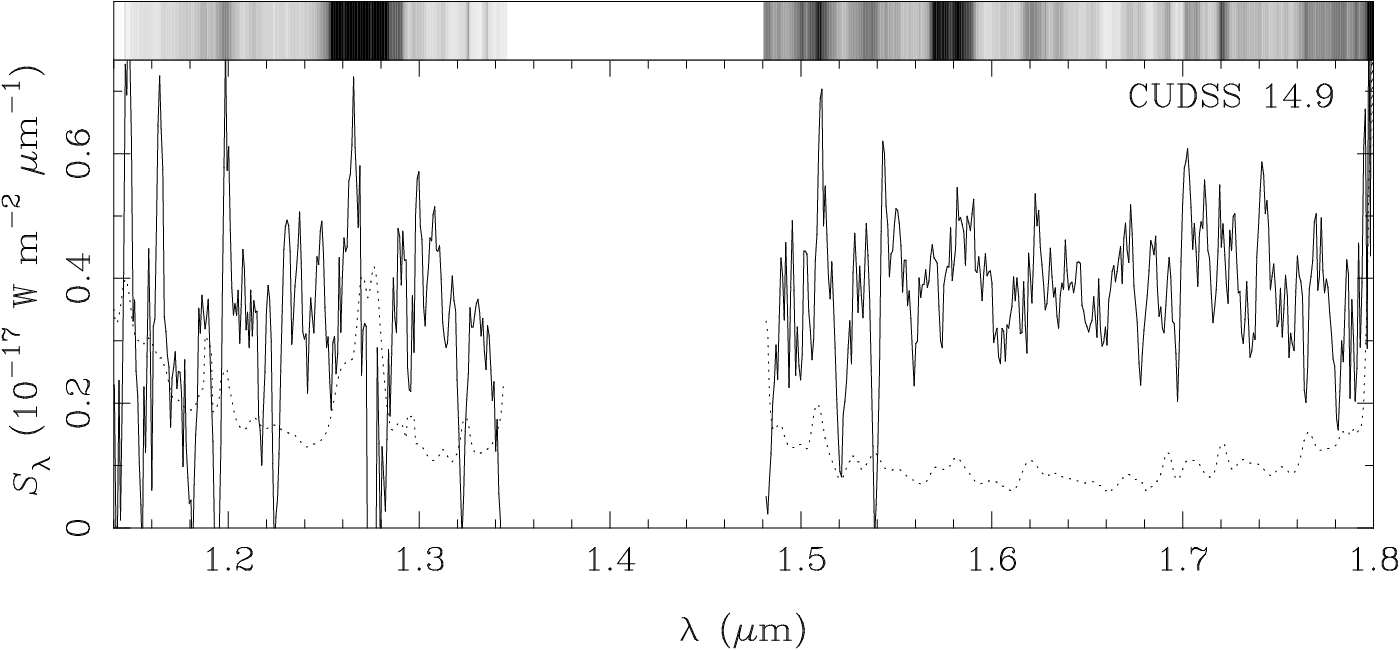}
\caption[]{Spectra of the targets, extracted in 1\,arcsecond wide
apertures. The solid line is the target spectrum, smoothed with a
5-pixel boxcar filter, while the dotted line indicates the $1\sigma$
uncertainty in the smoothed spectrum. The greyscale at the top
indicates the OH-suppressed sky brightness -- regions of high sky
brightness are subject to additional noise from imperfect sky
subtraction, which is not accounted for in the plotted noise
spectrum. The brightest sky features (black in the greyscale) are
multiplet O$_2$ transitions, which are not rejected by the mask.
Emission lines are labelled in those objects for which we have derived
a redshift.}
\label{fig:spectra}
\end{figure}

\addtocounter{figure}{-1}
\begin{figure}
\includegraphics[angle=-90,width=\colwidth]{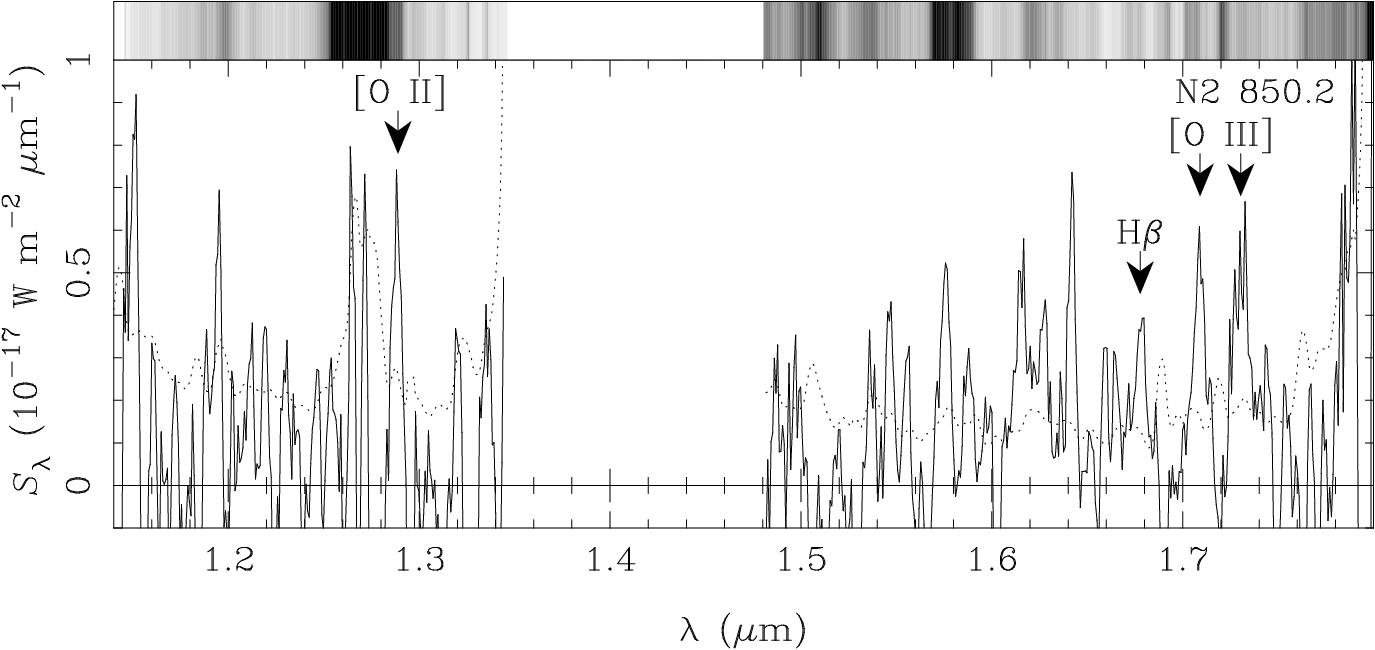}
\includegraphics[angle=-90,width=\colwidth]{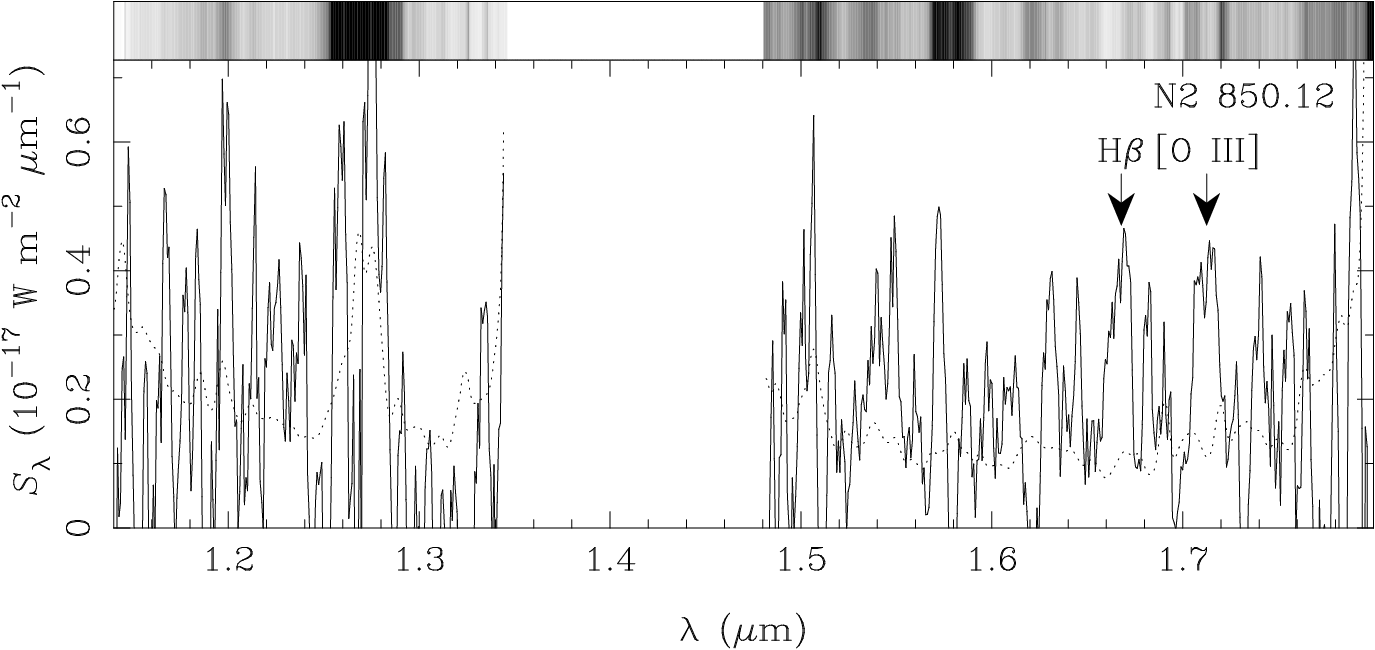}
\caption[]{continued}
\end{figure}

\section{Results}

The {\sc fwhm} of the arc lamp and planetary nebula lines was measured to
be 35--40\,\AA, indicating that the object spectra will have a spectral
resolution of $R \approx 200$. This is consistent with directly-made
measurements of the resolution obtained with a 1-arcsec slit (e.g.\ Willott
et al.\ 2003), and means that any emission lines will be unresolved. The
low resolution of the spectrum allows us to reject as spurious some
features which might at first glance appear real (especially in the
smoothed spectra presented in Fig.~\ref{fig:spectra}).  This is important
because the large number of pixels means that we have a $\sim 50$\,per cent
chance of finding a $3\sigma$ `emission line' simply by chance. This
probability is reduced by a factor of about 2 by excluding those `lines'
where the flux is dominated by a small number of errant pixels. On the
other hand, it is almost inevitable that each spectrum will contain at
least one emission feature which is formally significant at the $2.5\sigma$
level, and we must treat such marginal `detections' with extreme caution.

We can therefore only consider redshifts to be plausible if they are
determined from a single, highly significant line, or from two or more less
significant detections whose relative wavelengths correspond to a plausible
pair of features. To detect such features, we use a 51-pixel running median
to estimate the continuum level, and then search for groups of 9 pixels
(i.e.\ $\sim 75$\,\AA, or approximately the resolution FWHM) whose total
deviation from this continuum has a signal-to-noise ratio larger than a
certain threshold, which was set at 2.2 after some experimentation. Once
features had been detected in this manner, the 2D spectrum was examined to
ensure that they were not the result of poor sky subtraction or low-level
cosmic rays which had not been removed by the rejection algorithm. Not all
detections are listed therefore in Table~\ref{tab:lines}, where we
summarize the results of our analysis, although we only exclude those
features which are clearly spurious. The detections are described in detail
on an object-by-object basis in the next section. It is important to note
that the uncertainties quoted for the line fluxes are derived solely from
the detector read noise and photon shot noise, and do not take into
consideration the effects of improper sky subtractor or the placement of
the continuum. Incorrectly determining the continuum by as little as
$0.1\times10^{-17}\rm\,W\,m^{-2}\,\mu m^{-1}$ (see Fig.~\ref{fig:spectra})
will cause the measured line flux to be in error by $\sim
10^{-20}\rm\,W\,m^{-2}$, which is typically 20--40\,per cent of the
flux. We therefore warn against any interpretation of line ratios, which
are often substantially different from those expected. Even though in some
cases the median appears by eye to be wrong (and will always be slightly
biased by the presence of the detected line), any `by eye' estimate is
likely to be just as wrong, so we choose to fix the continuum at the level
derived by the running median, which at least has the advantage of being
non-subjective. The line centres are determined from a simple centroiding
of those pixels above the continuum level, and typically have an accuracy
of $\sim10$--20\,\AA\ which, when combined with the residuals from the
wavelength solution fit, imply uncertainties of $\pm0.005$ in each
individual redshift measurement.

Given the marginal nature of our line detections, we investigate the
probability of obtaining false redshifts from our data. We first
produce 100 artificial spectra from each of the ten noise spectra
(seven objects, plus three companions), by multiplying the noise
spectrum by a spectrum of random numbers drawn from a Normal
distribution. We apply our line detection algorithm to each of these
spectra, again rejecting those detections which are not consistent
with unresolved lines. Using the $2.2\sigma$ detection threshold, we
find that most spectra produce one or two spurious features, but there
is a pronounced tail, and 3\,per cent of our spectra produce five
features. While this may seem like cause for alarm, the fact that we
require more than half of the detected features to correspond to
plausible lines (and be absorption or emission features as
appropriate) makes it extremely unlikely that we will derive a
redshift from five randomly-located lines.

We investigate two distinct ways in which we can derive an erroneous
redshift. First, the redshift can be produced from entirely spurious
lines. This is not possible unless there are at least two lines, and
in this case the redshifts derived from their identifications must
agree to within 0.015 (for the case of three spurious lines, we
require that the redshifts from any pair agree, while for the cases of
four or five spurious lines, we require that consistent redshifts are
derived from a triplet). These probabilities can be calculated
directly from our 1000 spectra, and amount to 10\,per cent. Second, an
erroneous redshift can be produced from a single real line, plus one
or more spurious lines. To calculate this probability requires
\textit{a priori\/} knowledge of the probability of there being a
genuine feature at a given wavelength. If we assume that the genuine
emission line has an equal probability of appearing at any point in
the spectrum, then the probability of an erroneous redshift is 20\,per
cent, multiplied by the \textit{a priori\/} probability that there is
a genuine line in the $J$ or $H$ bands. The most likely cause of an
erroneous redshift arises from cases where three lines are detected,
and a pair of them correspond to plausible features.

\begin{figure}
\includegraphics[angle=-90,width=\colwidth]{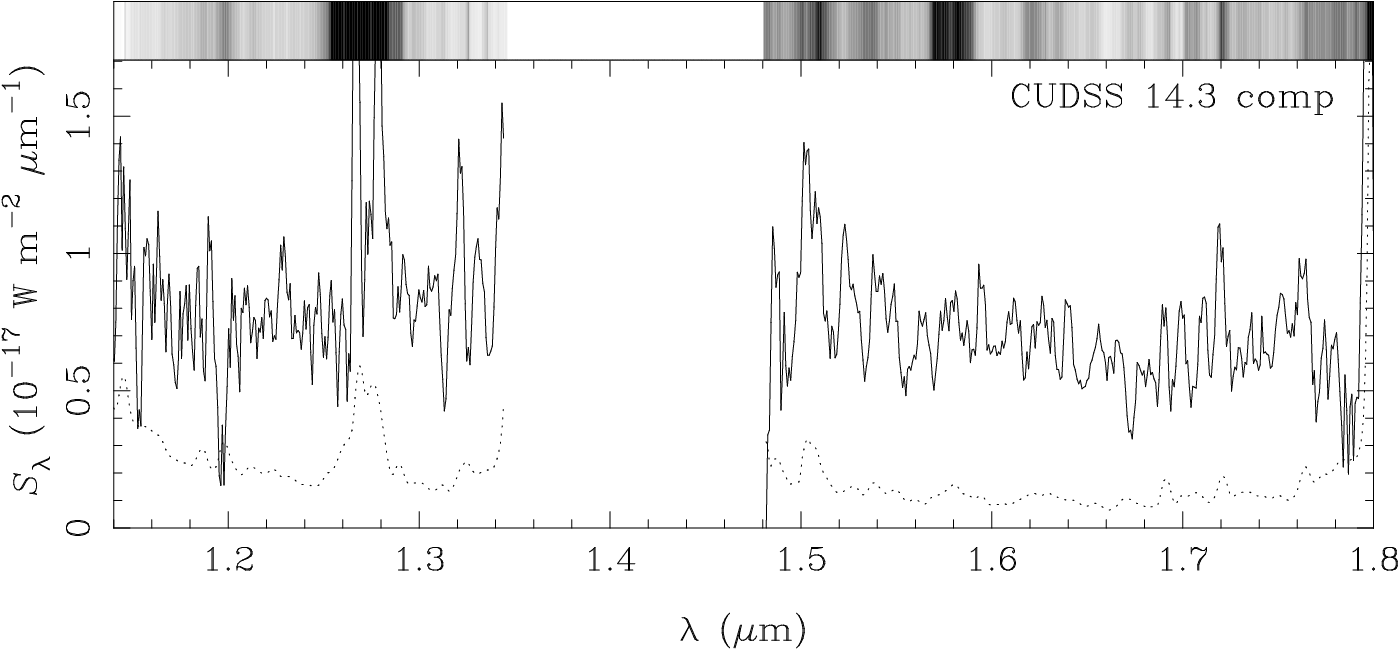}
\includegraphics[angle=-90,width=\colwidth]{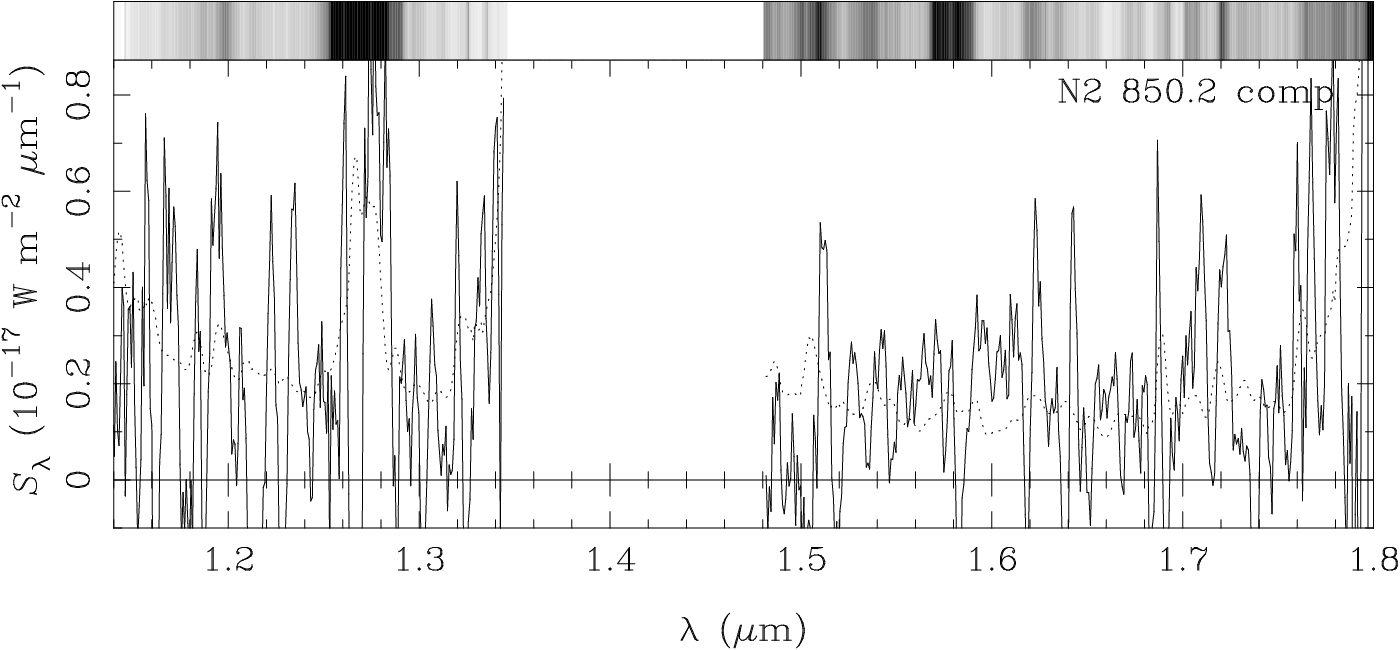}
\includegraphics[angle=-90,width=\colwidth]{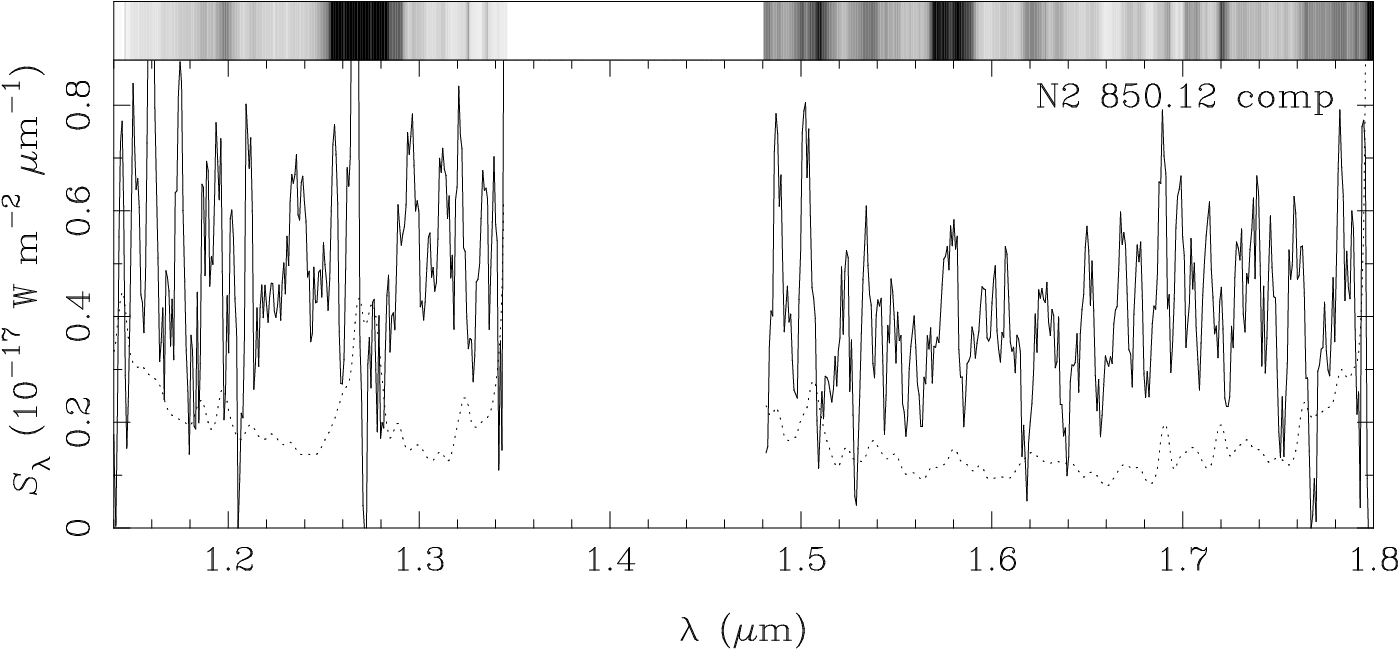}
\caption[]{Spectra of `companion' objects, presented in an identical manner
to the targets of Fig.~\ref{fig:spectra}.}
\label{fig:comp_spectra}
\end{figure}

\begin{table}
%\begin{minipage}{\textwidth}
\centering
\caption{Results of our line detection algorithm.\label{tab:lines}}
\begin{tabular}{lcrlc}
\hline
 & $\lambda_{\rm obs}$ & \multicolumn{1}{c}{Flux} \\
Object & ($\mu m$) & ($10^{-20}\rm\,W\,m^{-2}$) & Identification & $z$
\\
\hline
LE~850.1
& 1.1544 & $4.3\pm1.8$ & \\
& 1.2440 & $2.0\pm0.9$ & \\
& 1.5998 & $1.8\pm0.8$ & \\
\\
LE~850.3
& 1.1627 & $3.0\pm1.4$ & [O{\sc~ii}]~$\lambda$3727 & 2.120 \\
& 1.5993 & $1.6\pm0.7$ & spurious \\
\\
CUDSS~14.1
& 1.2994 & $1.9\pm0.8$ & & \\
& 1.6675 &$-1.6\pm0.6$ & \multicolumn{2}{c}{See text} \\
& 1.7227 & $2.3\pm1.0$ & \multicolumn{2}{c}{for discussion} \\
& 1.7543 & $2.1\pm0.7$ & \\
\\
CUDSS~14.3 & \multicolumn{4}{c}{Nothing detected} \\
\\
CUDSS~14.9
& 1.5216 &$-1.6\pm0.6$ & Ca~K? & 2.869 \\
& 1.5374 &$-2.0\pm0.8$ & Ca~H? & 2.874 \\
\\
N2~850.2
& 1.2889 & $4.0\pm1.8$ & [O{\sc~ii}]~$\lambda$3727  & 2.458 \\
& 1.5637 &$-2.4\pm0.8$ & spurious \\
& 1.5751 & $2.6\pm1.2$ & spurious \\
& 1.6777 & $2.1\pm0.9$ & H$\beta$ & 2.451 \\
& 1.7091 & $3.4\pm1.2$ & [O{\sc~iii}]~$\lambda$4959 & 2.446 \\
& 1.7304 & $3.9\pm1.3$ & [O{\sc~iii}]~$\lambda$5007 & 2.456 \\
\\
N2~850.12
& 1.5722 & $2.0\pm0.8$ & spurious \\
& 1.6678 & $2.2\pm0.8$ & H$\beta$ & 2.431 \\
& 1.7125 & $2.5\pm0.8$ & [O{\sc~iii}]~$\lambda$5007 & 2.420
\\
\hline
\multicolumn{5}{c}{Companion objects} \\
\hline
CUDSS~14.3 & \multicolumn{4}{c}{Nothing detected} \\
\\
N2~850.2 & 1.1700 & $3.8\pm1.6$ & \\
\\
N2~850.12 & \multicolumn{4}{c}{Nothing detected}
\\
\hline
\end{tabular}
%\end{minipage}
\end{table}

\section{Notes on individual objects}

In this section, we describe our redshift determinations based on the data
in Table~\ref{tab:lines}. We list all plausible emission lines, even if
they do not provide a definitive redshift measurement, so that our data may
be of use if an independent redshift determination supports the reality of
these lines. We deliberately refrain from making subjective comments on the
relative believability of any lines in a given spectrum.

\subsection{LE~850.1}

Due to the extremely faint and diffuse nature of this source, it took
a substantial amount of imaging on the night of UT 2002 May 21 to make
a firm imaging detection. Although by the time we had done so, the
hour angle of the source was no longer favourable, we confirmed the
offsets from a nearby star as being 11.6\,arcsec west and 5.2\,arcsec
south. We returned to LE~850.1 two nights later, and used our previous
imaging to hasten the acquisition. Since the distance between the
target and offset star is larger than the OHS field of view, we first
located the bright star near the end of the field of view, and made a
small (2\,arcsec) movement of the telescope by altering the location
of the centroiding box on the autoguider camera. The success of this
technique depends on the accuracy to which the plate scales of CISCO
and the AG camera are known, and we estimate the error to be no more
than 0.2\,arcsec. Since we were unable to obtain reliable photometry
from our acquisition image, we applied the same flux scaling factor
for LE~850.1 as was applied for LE~850.3; from the flux in the `clean'
part of our $H$-band spectrum (1.60--1.70\,$\mu$m), we estimate
$H=23.1\pm0.4$ in our extraction aperture.

Three emission lines are suggested by our line-finding algorithm, although
all are extremely marginal. No pair of these lines corresponds to plausible
features, and so we must conclude that at least two, and possible all
three, of the lines are spurious. However, given that it is three times
more likely for there to be only two spurious lines, rather than three, we
consider it probably that one of the lines is real. The redshift estimate
of $z=2.6^{+0.4}_{-0.5}$ (model `le2' of Aretxaga et al.\ 2003; hereafter
A03) suggests that either of the bluest two lines could be [O{\sc~ii}] (at
$z=2.097$ or $z=2.338$), but we are unable to assign a redshift from our
data.

\subsection{LE~850.3}

\begin{figure}
\includegraphics[angle=-90,width=\colwidth]{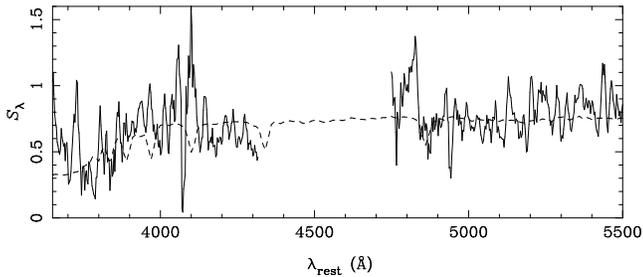}
\caption[]{Spectrum of LE~850.3 (as Fig.~\ref{fig:spectra}), overlaid
(dashed line) with the spectrum of a 250\,Myr-old starburst with an
$e$-folding time of 100\,Myr, reddened by $A_V=2.2$\,mag.}
\label{fig:le850p3_fit}
\end{figure}

This source is the brightest of our targets in the $H$-band ($H=20.4\pm0.2$ in
our extraction aperture) and has relatively good signal-to-noise in the clean
parts of the spectrum. Our detection algorithm produces two emission lines,
both at marginal ($<2.5\sigma$) significance.  The lines do not correspond to
any plausible pair of transitions, but we note the presence of a clear
continuum break at $\sim1.2\,\mu$m. We confirm the reality of this break by
subtracting the spectrum of blank sky at a distance of 1.5\,arcsec along the
slit (extracted in an identical 1-arcsec aperture). If we identify this break
as the Balmer jump, then the bluer of the two emission lines is [O{\sc~ii}] at
$z=2.120\pm0.004$, consistent with the radio--submm estimate of I02).

Fig.~\ref{fig:le850p3_fit} shows the near-IR spectrum of LE~850.3
(shifted to the rest frame), overlaid with a synthetic starburst model
(from the PEGASE code of Fioc \& Rocca-Volmerange 1997). This model has an
decaying star-formation rate with an $e$-folding time of 100\,Myr and is
seen at an age of 250\,Myr. It has been reddened by $A_V=2.2$, according to
Pei's (1992) parametrization of the Milky Way extinction law. Although this
is not a fit, it provides a reasonable description of the observed
continuum. The presence of a strong Balmer jump requires that the
rest-frame 4000-\AA\ light be dominated by $\sim10^8$\,yr-old stars,
although this may be because younger stars are more heavily obscured
(rather than because the star-formation rate is decaying with time, as in
our simple model). The reasonable fit that this model provides to the
spectrum gives us confidence in assigning a redshift of $z=2.12$ to this
source, even though it is rather inconsistent with the estimate of
$z=3.0^{+0.0}_{-0.5}$ of A03.

\subsection{CUDSS~14.1}

This source (also known as CUDSS~14A) was the subject of intense study by
Gear et al.\ (2000). High-resolution \textit{Hubble Space Telescope\/} and
ground-based (Canada--France--Hawaii Telescope) adaptive optics images
reveal it to be a red, compact galaxy. Deep near-IR \textit{JHK\/}
spectroscopy by Gear et al.\ failed to detect any emission lines to a level
of 1--$2 \times 10^{-19}\rm\,W\,m^{-2}$. Our new spectroscopy pushes this
limit almost an order of magnitude fainter, but fails to find any
significant ($>3\sigma$) lines. We detect four features (three emission
lines, and one absorption line) of marginal significance, however.

The measured wavelengths of the bluest and reddest emission lines are
consistent with being H$\beta$ and H$\alpha$, respectively, at
$z=1.673$. In this interpretation, the third emission line and the
absorption line are artefacts. Alternatively, these two lines could be
[O{\sc~ii}] at $z=2.486$ and [O{\sc~iii}]~$\lambda$5007 at $z=2.504$;
although this redshift discrepancy is large, it is not implausible given
the low signal-to-noise ratios of the detections. With these
identifications, the third emission line could be $\lambda$4959 at
$z=2.474$, which is barely consistent with the redshift from the other two
lines, while the absorption feature is, again, spurious. The lower redshift
is consistent with both the photometric redshift estimate of $z=1.25\pm0.3$
and the radio--submm redshift estimate of $z=1.9\pm0.48$ (Clements et al.\
2004; hereafter C04). We note, however, that A03 suggest a much higher
redshift of $z=3.8^{+0.7}_{-0.8}$, which is just consistent (at the 90\%
level) with both the above sets of line identifications. If this redshift
is correct, then one of the $H$-band emission lines could be [O{\sc~ii}] at
$z\sim3.7$, while the remaining three features would be artefacts (possibly
the 1.2994-$\mu$m line could be Mg{\sc~ii}~$\lambda$2798 at $z=3.644$ if
the 1.7227-$\mu$ line is [O{\sc~ii}] at $z=3.622$, but Mg{\sc~ii} is not
expected unless there is an AGN present). Ultimately, we decide against
assigning a spectroscopic redshift to this object.

\subsection{CUDSS~14.3}

Apart from a spurious peak due to poor subtraction of the unsuppressed OH
lines in the range 1.50--1.52\,$\mu$m, there are no significant features in
the spectrum of this faint ($H=21.6\pm0.3$) object. A brighter, slightly
bluer object is located approximately 2\,arcsec west and slightly north of
CUDSS~14.3 (fig.~4 of Webb et al.\ 2003), whose spectrum is similarly
featureless (Fig.~\ref{fig:comp_spectra}). We are unable to obtain a
redshift for either object. We note that both the photometric and
radio--submm redshift estimates of C04 are $z=1.1\pm0.3$, at which redshift
no strong emission lines would lie within our spectral coverage (H$\alpha$
would fall between the $J$ and $H$ spectroscopic windows). A03 obtain a
much higher redshift estimate ($z=3.2^{+0.8}_{-1.0}$), which would place
the [O{\sc~ii}] line somewhere between the 1.27-$\mu$m O$_2$ night sky
emission and the long-wavelength end of our spectral coverage, and it could
therefore easily fall between the $J$ and $H$ windows, or near the blue end
of the $H$-band where the poor sky subtraction affects our sensitivity.

\subsection{CUDSS~14.9}

The spectrum of CUDSS~14.9 ($H=20.9\pm0.2$) bears a remarkable similarity
to that of LE~850.3, with an apparent break in the continuum around
$1.2\,\mu$m.  However, the spectrum is noisier and the break is weaker --
if we make an \textit{ad hoc\/} definition od the break strength as $B =
\int_{1.21\mu\rm m}^{1.25\mu\rm m} S_\lambda \,{\rm d}\lambda /
\int_{1.14\mu\rm m}^{1.18\mu\rm m} S_\lambda \,{\rm d}\lambda$, we measure 
$B = 1.08 \pm 0.37$ for CUDSS~14.9, compared to $B = 1.47 \pm 0.24$ for
LE~850.3 (or $1.85 \pm 0.39$ after removal of the [O{\sc~ii}] line).

Two lines are detected by our algorithm; absorption features at 1.5216 and
1.5374\,$\mu$m with observed equivalent widths of $\sim40$\,\AA\ and
$\sim50$\,\AA, respectively. The wavelengths of these lines are consistent
with the Ca H+K lines at $z\approx2.87$. We are hesitant about assigning
this redshift, however, since there is no clear continuum break associated
with the absorption lines (although there is insufficient coverage to the
blue of them to make an accurate measurement of the continuum level) and
the redder of the two features is affected by a night sky emission line.
Furthermore, this redshift is inconsistent with C04's photometric and
radio--submm estimates of $z=1.44\pm0.32$ and $z=1.7\pm0.43$, respectively,
although it agrees with A03's estimate of $z=2.5^{+0.5}_{-1.0}$. If our
spectroscopic measurement is correct, however, none of the strong
rest-frame optical emission lines associated with star formation
(H$\alpha$, H$\beta$, [O{\sc~ii}]) would lie within a spectroscopic window.

\subsection{ELAIS~N2~850.2}

The identification for this source is extremely faint ($H=22.4\pm0.4$
in our spectroscopic aperture) and the continuum is only seen in our
2D spectrum after substantial smoothing. The location of the
extraction apertures were determined from the acquistion image. Five
plausible emission features (one of which is detected at $>3\sigma$)
and one absorption feature are detected by our algorithm, although the
absorption feature is close to one of the emission lines and may
result from imperfect subtraction of the 1.58-$\mu$m sky emission. The
other four emission lines, however, are consistent with a redshift
$z=2.453\pm0.006$. This redshift is consistent with tha lack of
formally-detected continuum in the $J$-band, since the 4000-\AA\ break
will lie between the $J$ and $H$ atmospheric windows, and also with
the redshift estimates of I02 and A03. It also agrees well with the
optical redshift of 2.454 (Chapman et al., in preparation).

Our spectroscopic slit was aligned to pass through a galaxy 2\,arcsec
south--southeast of ELAIS~N2~850.2 (fig.~2 of I02). This source shows faint
continuum in the clean part of the $J$-band ($1.20 < \lambda < 1.26\,\mu$m;
the continuum is more clearly seen in the smoothed 2D spectrum than in
Fig.~\ref{fig:comp_spectra}) but this does not rule out its being at the
same redshift as the submm source, since it is brighter ($H=22.1\pm0.3$ in
our aperture). There is a single marginally-detected emission line at
1.1700\,$\mu$m; if real, this would mean that the `companion' is not
physically associated with the submm source, since it does not correspond
to any known emission line at the redshift of the submm source.

\subsection{ELAIS~N2~850.12}

This identification for this source is also very faint ($H=22.2\pm0.4$) and
the signal-to-noise ratio of our spectrum is low. However, we find three
emission lines, of which we identify two as H$\beta$ and
[O{\sc~iii}]~$\lambda$5007 at $z = 2.425 \pm 0.005$ (the absence of
detectable [O{\sc~iii}]~$\lambda$4959 is simply due to its low flux). This
redshift is also consistent with the lack of continuum at the red end of
the $J$-band, since it places the 4000-\AA\ break at $1.37\,\mu$m (we note,
however, that continuum is detected at the $\sim 3\sigma$ level in the
`clean' part of the $J$-band (1.20--1.25\,$\mu$m). At this redshift, the
[O{\sc~ii}]~$\lambda$3727 emission line would lie at $1.276\pm0.001\,\mu$m,
where our sensitivity is low ($3\sigma \approx 7 \times
10^{-20}\rm\,W\,m^{-2}$) due to the O$_2$ night-sky emission; its
non-detection is therefore not at odds with a ratio [O{\sc~ii}]/H$\beta
\approx 3$ (e.g.\ Kennicutt 1998), especially if there is some
reddening. We note that this redshift is unexpectedly low, given the
high ratio of submm to radio flux, though this inconsistency can be
understood if the handful of 3$\sigma$ radio peaks noted by I02 are
the relics of a diffuse radio source. Since this redshift is derived
from two of three detected lines, it should be treated with caution,
although it gains some support from the fact that there is positive
flux at the expected position of [O{\sc~ii}] (formally, the value is
$(3.8\pm2.2) \times 10^{-20}\rm\,W\,m^{-2}$), and from the relatively
high significance of the line at 1.7125\,$\mu$m (${\rm S/N}>3$).

Our spectroscopic slit was aligned to pass through a second object located
approximately 7\,arcsec west of N2~850.12 (on the edge of the postage stamp
shown in fig.~2 of I02). This object is brighter ($H=20.2\pm0.1$) and bluer
and our algorithm failed to detect any emission lines in its spectrum (see
Fig.~\ref{fig:comp_spectra}). We cannot say whether it is physically
associated with the submm galaxy.

\section{On the absence of emission lines}

\begin{figure}
\includegraphics[angle=-90,width=\colwidth]{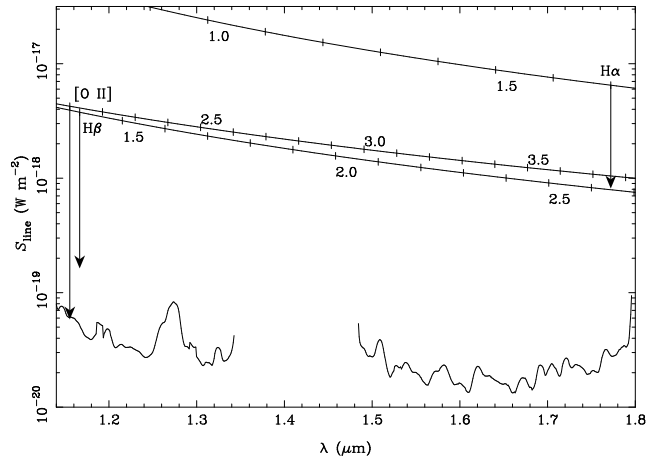}
\caption[]{Sensitivity to emission lines associated with star formation.
The bottom line indicates our typical $3\sigma$ sensitivity to an
unresolved emission line as a function of wavelength. The three curves near
the top of the plot show the wavelengths and line fluxes expected for a
$\rm1000\,M_\odot\,yr^{-1}$ starburst (Kennicutt 1992, 1998). Vertical tick
marks are placed at redshift intervals of 0.1, and are numbered at
intervals of 0.5 Finally, the downward-pointing vertical arrows indicate
the effect of $A_V=3$\,mag of internal reddening.
\label{fig:lineflux}}
\end{figure}

Given the extreme star-formation rates implied by the submm luminosities of
these objects and our extensive wavelength coverage, the lack of strong
emission lines is perhaps a surprise. A star-formation rate of
$\rm1000\,M_\odot\,yr^{-1}$ (typical of these objects; Ivison et al.\ 1998;
Gear et al.\ 2000) corresponds to an H$\alpha$ luminosity $L_{\rm H\alpha} =
1.3 \times 10^{37}$\,W (Kennicutt 1998), and we assume H$\alpha$/H$\beta =
2.78$ (Case~B; e.g.\ Osterbrock 1989) and H$\alpha$/[O{\sc~ii}$] = 0.9$
(Kennicutt 1992). These lines should all be well above our detection threshold
(see Fig.~\ref{fig:lineflux}). In fact, in the absence of any extinction, our
2-hour integrations should be sensitive to star-formation rates as low as $\sim
20\rm\,M_\odot\,yr^{-1}$ out to $z \approx 3.5$, although there is a gap in our
coverage between $2.6 \la z \la 3.0$ when [O{\sc~ii}] lies in the gap between
the $J$ and $H$ atmospheric windows. This gap could be significant since it
lies close to the peak in the redshift distribution, $z_{\rm median} \sim 2.4$,
presented by Chapman et al.\ (2003), based on optical spectroscopy of bluer
radio-detected submm galaxies.

We have redshifts for only three of our seven targets. Taken at face
value, the measured line fluxes of LE~850.3, N2~850.2, and N2~850.12
correspond to star-formation rates of $\sim10$, $\sim20$, and
$\sim25\rm\,M_\odot\,yr^{-1}$, respectively, well below the estimates from
the radio--submm data (I02). Such large discrepancies are not unusual; for
example, in the extremely red object HR~10 (Hu \& Ridgway 1994), the raw
(i.e.\ uncorrected for extinction) H$\alpha$-derived star-formation rate is
a factor of $\sim 30$ lower than that derived from the submm flux (Cimatti
et al.\ 1998; Dey et al.\ 1999). The two rates can be reconciled by
appealing to substantial quantities of dust which severely attenuate the
observed line emisison.  In all three cases, the emission-line-derived
star-formation rate can be rectified with the submm--radio-derived SFR if
the emission lines are extinguished by $A_V \approx 3$--4\,mag (compared to
the $A_V \approx 4.5$\,mag needed in HR~10\footnote{Dey et al.\ (1999)
claim that a simple foreground dust screen is inapproriate because it would
imply an intrinsic ratio of [O{\sc~ii}]/H$\alpha \approx 1.3$, which they
claim is ``at the extreme limit of observed values for local star-forming
galaxies (e.g., Kennicutt 1992).'' However, Fig.~12 of Kennicutt (1992)
indicates that this ratio is quite normal, since the data plotted are
uncorrected for reddening (a correction factor of $\sim 2$) or [N{\sc~ii}]
contamination ($\sim 1.5$).}). Slit losses are unlikely to account for more
than a factor of about 2 in our line fluxes since our slit width
corresponds to a linear size of 8\,kpc, which is large for a starburst
region (Smail et al.\ 2003). Furthermore, the radio emission is not
observed to be extended on scales much larger than this.

Arguing against a foreground screen hypothesis is the detection of
Ly$\alpha$ emission in submillimetre galaxies (Chapman et al.\ 2003). The
measured fluxes correspond to star formation rates of
$\sim10\rm\,M_\odot\,yr^{-1}$, broadly in line with our measurements, while
a foreground screen with $A_V=3$ would produce at least four orders of
magnitude of extinction at Ly$\alpha$ and completely suppress the line. The
consistency between the UV and optical star formation diagnostic lines
therefore suggests that we are seeing a small fraction ($\sim 1$ per cent)
of the star formation in a relatively unobscured way, while the vast
majority is completely extinguished at optical/UV wavelengths. Some
obscuration is still required since a $10\rm\,M_\odot\,yr^{-1}$ SFR at
$z=2.4$ produces an observed $V\approx22$ (Leitherer, Robert, \& Heckman
1995), which is brighter than our targets (I02; C04). This picture is quite
reasonable, since massive stars are formed in the local Universe inside the
cores of dense molecular clouds, and the $\sim 15$\,Myr it takes for these
stars to disperse their parent clouds is longer than the lifetimes of stars
with $M \ga 15$--20\,M$_\odot$ (see discussion in Jimenez et al.\ 2000).
Jimenez et al.\ suggest that the observed flux of far-ultraviolet photons
represents only about one-sixth of the total star formation, and the
observed line emission should also underestimate the true star formation
rate by a similar factor if the massive stars are heavily obscured. Larger
discrepancies are possible if the starburst is younger than 15\,Myr since
even the longer-lived stars are embedded in their parent clouds.

An alternative hypothesis which warrants investigation is that the star
formation rates derived from the submm observations are substantially
overestimated. The most likely reason for this would be erroneously high
assumed dust temperatures. For a dust emissivity index $\beta$, the
bolometric luminosity per unit dust mass is proportional to $T^{4+\beta}$,
while the dust mass for a given monochromatic submm luminosity is inversely
proportional to the dust temperature. Therefore, for a source with a known
redshift and measured 850-$\mu$m flux, the inferred bolometric dust
luminosity (and hence star formation rate) is proportional to $T^{3+\beta}$
(e.g.\ Blain et al.\ 2002). This suggests that very large errors could be
introduced by an incorrect estimate of the dust temperature. However, the
expression is only accurate at low redshift, where $h\nu_{\rm e} \ll kT$
($\nu_{\rm e}$ being the frequency at which the observed 850-$\mu$m
radiation is emitted), and the possible error is more modest at $z \sim 2$,
where a factor of 2 uncertainty in dust temperature corresponds to a factor
of 10 uncertainty in luminosity (e.g.\ Hughes, Dunlop \& Rawlings 1997;
Eales et al.\ 2000).

I02 adopt $T_{\rm dust} = 45$\,K, which is near the hot end of the
distribution found for ultraluminous infrared galaxies (Dunne et al.\ 2000;
Farrah et al.\ 2002). In principle, therefore, the star formation rates of
$\sim1000\rm\,M_\odot\,yr^{-1}$ could be an order of magnitude too high if
the temperature of the dust is $\sim20\rm\,K$. The ratio of 850-$\mu$m to
450-$\mu$m flux is a relatively sensitive function of dust temperature for
$z \ga 1$, as the shorter wavelength point approaches the peak of the
emission. Eales et al.\ (2000) claim an average upper limit of
$S_{850}/S_{450} \ga 0.5$ for the CUDSS sources, which appears to rule out
$T\sim40\rm\,K$ dust if the median source redshift is 2. However, Fox et
al.\ (2002) only constrain this ratio to be $\ga 0.3$ for sources from the
8-mJy survey, which is low enough to allow hotter dust. In principle, the
radio flux should provide an alternative estimate of the SFR if the
redshift is known, since the tight radio--FIR correlation provides a way to
derive the total FIR luminosity without having to assume the dust
temperature. Systematically overestimated SFRs should result in high values
of $S_{850\rm\,\mu m}/S_{1.4\rm\,GHz}$ (i.e.\ steeper spectral indices)
whereas there appears to be a strong tendency for the observations to
produce shallower spectra (e.g.\ C04). The cause of this is unclear, and it
seems that this issue will only be unequivocally resolved with shorter
wavelength detections which can constrain the shape of the SED near the
peak of the thermal emission. For now, we believe we are justified in
claiming that it is highly improbable that the submm-derived SFRs are
reconcilable with our spectroscopically-derived estimates unless most of
the star formation is heavily obscured. We are unable to determine whether
this obscured fraction is 90\,per cent or 99\,per cent of the total star
formation.

\section{Comparison with other spectroscopic studies and future prospects}

Near-infrared spectroscopy of submm galaxies has been rather limited
to date. Ivison et al.\ (2000) presented $K$-band spectra of two
sources behind the lensing cluster Abell~1835, SMM~J14011+0252 and
SMM~J14010+0253, which both showed H$\alpha$ emission but lacked the
strong [N{\sc~ii}] emission characteristic of an AGN. Using
near-infrared spectrographs on larger telescopes, Frayer et al.\
(2003) presented a spectrum of SMM~J04431+0210, while Smail et al.\
(2003a,b) observed SMM~J17142+5016 (in the protocluster around
LBDS~53W002) and ELAIS~N2~850.4, respectively. All three of these
galaxies showed strong emission lines indicative of the presence of an
AGN, which Smail et al.\ suggest is a common feature of these
objects. All the emission lines in these objects are brighter than our
upper limits, by an order of magnitude in the cases of SMM~J17142+5016
and N2~850.4. The spectra of our targets are dramatically different
from those of the (few) objects in the literature. However, this fact
may be readily explicable by the fact that our targets were selected
to be optically faint, and are therefore perhaps biased towards more
heavily extinguished objects. On the other hand, previously observed
objects already had known redshifts from optical spectroscopy, and
must therefore have had bright Ly$\alpha$ emission, and/or
sufficiently bright optical continua to allow the detection of
absorption lines. Either case implies that the objects with near-IR
spectra in the literature are lightly extinguished.

In support of this idea, Eisenhauer et al.\ (2003) recently obtained a
high-quality near-infrared spectrum of the $z=2.565$ submm source
SMM~J14011+0252. They found that the rest-frame optical continuum of
this object is dominated by stars with ages of a few hundred Myr,
similar to our estimate for LE~850.3 (the only object in which we
detect an age-sensitive stellar feature) although, unlike LE~850.3,
the continuum of SMM~J14011+0252 appears to be unreddened, which
presumably explains its unusually bright rest-frame UV spectrum
(Ivison et al.\ 2001). Such an age is perhaps larger than expected,
since a constant star formation rate of
$\sim1000\rm\,M_\odot\,yr^{-1}$ can form a massive galaxy in only
$\sim100\rm\,Myr$.

The most serious difficulty which arises when attempting to understand
the nature of submm galaxies as a class comes from a lack of clarity
in the criteria with which submm galaxies are selected for
spectroscopy by different groups. This difficulty is almost certainly
compounded by the likelihood that spectra are only published where
firm redshifts are obtained (i.e., objects with luminous emission
lines, or those with bright enough continua to allow the detection of
absorption features). This leads to the presence of strong and
uncertain selection effects in the literature samples, and makes
generalized statements about the properties of the population as a
whole unreliable.

Nevertheless, almost 100 spectroscopic redshifts for submm galaxies have
been obtained, and there is little reason to believe that the redshift
distribution obtained from these objects is significantly different from
that of the class as a whole (Chapman et al.\ 2003; but see also
Dannerbauer et al.\ 2004). It is therefore clearly time for the study of
these important objects to progress from simple redshift determination and
basic classification (AGN, starbust, composite) to more detailed
analyses. The difficulties posed by the presence of large quantities of
dust and the existence of multiple components point to the need for
spectroscopy over a long wavelength baseline, covering rest-frame
wavelengths from Ly$\alpha$ to H$\alpha$.

\begin{figure}
\includegraphics[angle=-90,width=\colwidth]{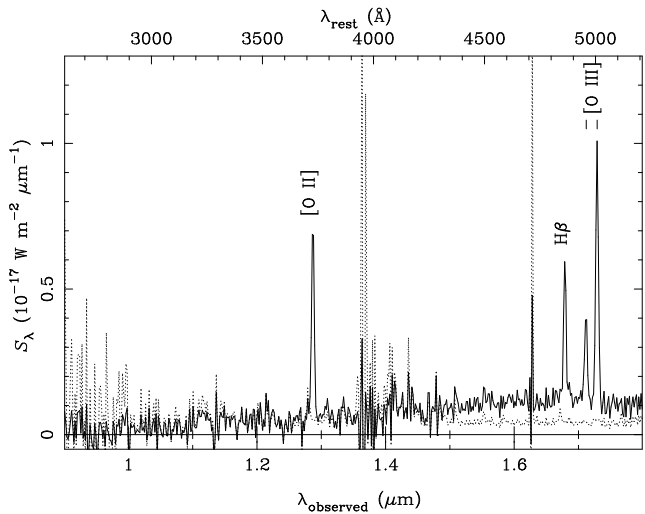}
\caption[]{Simulated 7-hour FMOS spectrum of ELAIS~N2~850.2. The spike at
1.62\,$\mu$m is caused by the masking of multiple airglow lines which
effectively remove all the flux from that pixel, and is not
significant (as indicated by the dotted line which shows the $1\sigma$
noise). See text for more details.}
\label{fig:fmos}
\end{figure}

The most interesting new instrument for this purpose is FMOS, the
Fibre Multi-Object Spectrograph (Maihara et al.\ 2000; Kimura et al.\
2003) which is scheduled to be commissioned on Subaru Telescope in
late 2005. This instrument will provide OH-suppressed spectroscopy
with increased sensitivity, spectral resolution, and wavelength
coverage (0.9--1.8$\,\mu$m) compared to OHS, using 400 fibres across a
30-arcminute field of view (this corresponds to the source density at
a flux level of $S_{850} \ga 3\rm\,mJy$; e.g., Borys et al.\
2003). The ability to take many spectra simultaneously enables much
longer exposures than we were able to undertake here and, coupled with
the increased sensitivity of the new instrument, it should be
relatively straightforward to probe a further order of magnitude below
our current detection limits. Our emission line sensitivities will
therefore correspond to star formation rates of $\sim
1\rm\,M_\odot\,yr^{-1}$, and it would be remarkable if a significant
number of targets failed to display emission lines. Furthermore, the
increased sensitivity to the continuum emission will allow more
accurate modelling of the galaxies' stellar populations. As a
demonstration of the likely data quality we can expect from FMOS, we
simulated the spectrum of N2~850.2 using the FMOS Spectral Simulator
developed by Naoyuki Tamura\footnote{The FMOS Spectrum Simulator can
be found at
http://elvira.phyaig.dur.ac.uk/naoyuki.tamura/simulator.html}
(Fig~\ref{fig:fmos}). Our input spectrum assumes fluxes for the
H$\beta$, [O{\sc~ii}], and [O{\sc~iii}]~$\lambda$5007 lines given in
Table~\ref{tab:lines}, superimposed on a continuum of a 3-Gyr-old
elliptical, reddened by $E(B-V)=0.5$. All emission lines are assumed
to have $\rm FWHM = 500\,km\,s^{-1}$, and an exposure time of 7\,hours
was adopted, representing the maximum likely on-source integration
time possible in a single night's observing.

A major caveat to this lies in the infrared ``spectroscopic desert''
at $2.6 \la z \la 3.0$. Here, [O{\sc~ii}], the Balmer jump, and the
4000-\AA\ break are redshifted between the $J$ and $H$ windows;
H$\beta$ and [O{\sc~iii}] are redshifted between $H$ and $K$; and
H$\alpha$ is redshifted beyond the $K$ band. Although this is a
relatively small redshift range, it is close to the peak of the
redshift distribution and may include more than 10 per cent of sources
from a flux-limited sample (Chapman et al.\ 2003). Without the
reddening-insensitive age measurements provided by the Balmer jump
and/or 4000-\AA\ break, and with the low spectral resolution of FMOS,
detailed stellar population analysis of these targets is unlikely to
prove fruitful.

For now, further infrared spectroscopy is still worthwhile as a precursor
to the above observations. Many submm sources display optical emission
lines and, regardless of whether these are powered by a starburst or an
AGN, there will be associated nebular lines redshifted into the
near-infrared. The relative strengths of these lines (in objects whose
redshifts are suitable for near-infrared spectroscopy) will test our idea
that they come from a relatively unobscured region comprising only a small
fraction of the total starburst.

\section{Summary}

We have presented the results of deep near-IR spectroscopy of seven
galaxies selected from the SCUBA 8-mJy and CUDSS submm surveys. We have
obtained spectroscopic redshifts for three of these objects, two of which
(LE~850.3 and ELAIS~N2~850.2) agree well with the prior estimates from
radio--submm data, while the other (ELAIS~N2~850.12) is unexpectedly
low. These redshifts are based on marginal ($\la3\sigma$) detections of
emission lines, although in all cases they come from more than one spectral
feature, and in one instance (N2~850.2) there are four emission lines, and
the derived redshift agrees with the (as yet unpublished) Ly$\alpha$ and CO
redshifts. In all three cases, the star-formation rates derived from the
emission-line luminosities are much lower than those derived from the
radio--submm data. By comparing the bulk properties of our sample with
those observed at optical wavelengths, we suggest that we are seeing a
small fraction of the total starburst through low extinction. Further
infrared spectroscopy of sources with secure optical redshifts are required
to test this hypothesis.

\section*{Acknowledgments}

This paper is based on data collected at Subaru Telescope, which is
operated by the National Astronomical Observatory of Japan. The
authors are grateful to the staff of Subaru Telescope for their help
with the observations, and to the anonymous referee for his/her
comments. CS and JSD thank the Particle Physics and Astronomy Research
Council for funding in the form of an Advanced Fellowship and Senior
Research Fellowship, respectively.

\bsp

\label{lastpage}


\begin{thebibliography}{99}

\bibitem{}Aretxaga I., Hughes D. H., Chapin E. L., Gazta\~{n}aga E., Dunlop
J. S., Ivison R. J., 2003, MNRAS, 342, 759 (A03)

\bibitem{}Blain A. W., Kneib J.-P, Ivison R. J., Smail I., 1999, ApJ,
512, L87

\bibitem{}Blain A. W., Smail I., Ivison R. J., Kneib J.-P., Frayer D. T.,
2002, Phys.\ Rep., 369, 111

\bibitem{}Borys C., Chapman S., Halpern M., Douglas S., 2003, MNRAS,
344, 385

\bibitem{}Carilli C. L., Yun M. S., 1999, ApJ, 513, L13

\bibitem{}Carilli C. L., Yun M. S., 2000, ApJ, 530, 618

\bibitem{}Chapman S. C., Blain A. W., Ivison R. J., Smail I., 2003, Nat,
422, 695
\bibitem{}Cimatti A., Andreani P., R\"{o}ttgering H., Tilanus R.,
1998, Nat, 392, 895

\bibitem{}Clements D., et al., 2004, MNRAS, in press (astro-ph/0312269)
(C04)

\bibitem{}Dannerbauer H., Lehnert M. D., Lutz D., Tacconi L., Bertoldi F.,
Carilli C., Genzel R., Menten K., 2004, ApJ, 573, 473

\bibitem{}Dey A., Graham J. R., Ivison R. J., Smail I., Wright G. S., Liu
M. C., 1999, ApJ, 519, 610

\bibitem{}Downes D., et al., 1999, A\&A, 347, 809

\bibitem{}Dunlop J. S., et al., 2004, MNRAS, in press (astro-ph/0205480)

\bibitem{}Dunne L., Clements D., Eales S. A., 2000, MNRAS, 319, 813

\bibitem{}Eales S., Rawlings S., 1993, ApJ, 411, 67

%\bibitem{}Eales S., Lilly S., Gear W., Dunne L., Bond J. R., Hammer F., Le
%F\`{e}vre O., Crampton, D., 1999, ApJ, 515, 518

\bibitem{}Eales S., Lilly S., Webb T., Dunne L., Gear W., Clements D., Yun
M., 2000, AJ, 120, 2244

\bibitem{}Eisenhauer F., et al., 2003, ESO Messenger, 113, 17

\bibitem{}Farrah D., Serjeant S., Efstathiou A., Rowan-Robinson M., Verma
A., 2002, MNRAS, 335, 1163

\bibitem{}Fioc M., Rocca-Volmerange B., 1997, A\&A, 326, 950

\bibitem{}Fox M. J., et al., 2002, MNRAS, 331, 839

\bibitem{}Frayer D. T., Armus L., Scoville N. Z., Blain A. W., Reddy
N. A., Ivison R. J., Smail I., 2003, AJ, 126, 73

\bibitem{}Gear W. K., Lilly, S. J., Stevens J. A., Clements D. L., Webb
T. M., Eales S. A., Dunne L., 2000, MNRAS, 316, L51

\bibitem{}Hauser M. G., et al., 1998, ApJ, 508, 25

\bibitem{}Hawarden T. G., Leggett S. K., Letawsky M. B., Ballantyne D. R.,
Casali M. M., 2001, MNRAS, 325, 563

\bibitem{}Hu E. M., Ridgway S. E., 1994, AJ, 107, 1303

\bibitem{}Hughes D. H., Dunlop J. S., Rawlings S., 1997, MNRAS, 289, 766

\bibitem{}Hughes D. H., et al., 1998, Nat, 394, 241

\bibitem{}Hughes D. H., et al., 2002, MNRAS, 335, 871

\bibitem{}Ivison R. J., Smail I., Le Borgne J.-F., Blain A. W., Kneib
J.-P., B\'{e}zecourt J., Kerr T. H., Davies J. K., 1998, MNRAS, 298, 583

\bibitem{}Ivison R. J., Smail I., Barger A. J., Kneib J.-P., Blain
A. W., Owen F. N., Kerr T. H., Cowie L. L., 2000, MNRAS, 315, 209

\bibitem{}Ivison R. J., Smail I., Frayer D. T., Kneib J.-P, Blain
A. W., 2001, ApJ, 561, L45

\bibitem{}Ivison R. J., et al., 2002, MNRAS, 337, 1 (I02)

\bibitem{}Iwamuro F., Motohara K., Maihara T., Hata R., Harashima T.,
2001, PASJ, 53, 355

\bibitem{}Jimenez R., Padoan P., Dunlop J. S., Bowen D. V., Juvela M.,
Matteucci F., 2000, ApJ, 532, 152

\bibitem{}Kennicutt R. C., 1992, ApJ, 388, 310

\bibitem{}Kennicutt R. C., 1998, ApJ, 498, 541

\bibitem{}Kimura M., et al., 2003, in Iye M., Moorwood A. F., eds,
Proc.\ SPIE 4841. SPIE, Bellingham, p.~974

\bibitem{}Leitherer C., Robert C., Heckman T. M., 1995, ApJS, 99, 173

\bibitem{}Maihara T., et al., 2000, in Iye M., Moorwood A. F., eds,
Proc.\ SPIE 4008. SPIE, Bellingham, p.~1111

\bibitem{}Motohara K., et al., 2002, PASJ, 54, 315

\bibitem{}Osterbrock D. E. 1989, Astrophysics of Gaseous Nebulae and
Active Galactic Nuclei (Mill Valley: University Science Books)

\bibitem{}Pei Y. C., 1992, ApJ, 395, 130

\bibitem{}Perryman M. A. C., et al., 1997, A\&A, 323, L49

\bibitem{}Rengarajan T. N., Takeuchi T. T., 2001, PASJ, 53, 433

\bibitem{}Scott S. E., et al., 2002, MNRAS, 331, 817

\bibitem{}Smail I., Ivison R. J., Gilbank D. G., Dunlop J. S., Keel
W. C., Motohara K., Stevens J. A., 2003a, ApJ, 583, 551

\bibitem{}Smail I., Chapman S. C., Ivison R. J., Blain A. W., Takata
T., Heckman T. M., Dunlop J. S., Sekiguchi K., 2003b, MNRAS, 342, 1185

\bibitem{}Steidel C. C., Pettini M., Hamilton D., 1995, AJ, 110, 2519

\bibitem{}Tokunaga A. T., 2000, in Cox A. N. ed., Allen's Astrophysical
Quantities (4th Edition). Springer, Berlin, p.~151

\bibitem{}Webb T. M. A., Lilly S. J., Clements D. L., Eales S., Yun M.,
Brodwin M., Dunne L., Gear W. K., 2003, ApJ, 597, 680

\bibitem{}Willott C. J., et al., 2003, MNRAS, 339, 397

\bibitem{}Yun M. S., Carilli C. L., 2002, ApJ, 568, 88

\end{thebibliography}
\end{document}